\documentclass[aps,prl,reprint,amsmath,amssymb,superscriptaddress,showpacs]{revtex4-1}
\usepackage{bm}
\usepackage{graphicx}
\usepackage{color}
\usepackage{multirow}

\begin{document}

\title{Frustrated antiferromagnetic honeycomb-tunnel-like lattice Cu\textit{RE}$_2$Ge$_2$O$_8$ (\textit{RE}=Pr, Nd, Sm, and Eu)}

\author{Hwanbeom Cho}
\affiliation{Center for Correlated Electron Systems, Institute for Basic Science (IBS), Seoul 08826, Republic of Korea}
\affiliation{Department of Physics and Astronomy, Seoul National University, Seoul 08826, Republic of Korea}

\author{Marie Kratochv\'ilov\'a}
\affiliation{Center for Correlated Electron Systems, Institute for Basic Science (IBS), Seoul 08826, Republic of Korea}
\affiliation{Department of Physics and Astronomy, Seoul National University, Seoul 08826, Republic of Korea}

\author{Nahyun Lee}
\affiliation{Center for Correlated Electron Systems, Institute for Basic Science (IBS), Seoul 08826, Republic of Korea}

\author{Hasung Sim}
\affiliation{Center for Correlated Electron Systems, Institute for Basic Science (IBS), Seoul 08826, Republic of Korea}
\affiliation{Department of Physics and Astronomy, Seoul National University, Seoul 08826, Republic of Korea}

\author{Je-Geun Park}
\email{jgpark10@snu.ac.kr}
\affiliation{Center for Correlated Electron Systems, Institute for Basic Science (IBS), Seoul 08826, Republic of Korea}
\affiliation{Department of Physics and Astronomy, Seoul National University, Seoul 08826, Republic of Korea}

\date{\today}

\begin{abstract}
New frustrated antiferromagnetic compounds Cu$RE_2$Ge$_2$O$_8$ ($RE$=Pr, Nd, Sm, Eu) have been investigated using high-resolution x-ray diffraction, magnetic and heat capacity measurements. These systems show different magnetic lattices depending on rare-earth element. The nonmagnetic Eu compound is a $S$=1/2 two-dimensional triangular antiferromagnetic lattice oriented in the $ac$ plane with geometrical frustration. On the other hand, the Pr, Nd, and Sm compounds show a three-dimensional honeycomb-tunnel-like lattice made of $RE^{3+}$ running along the $a$ axis with the characteristic behavior of frustrated antiferromagnets. \end{abstract}
\maketitle

\section{I. INTRODUCTION}
Geometrical frustration in magnetic materials has attributed significant attention over the last few decades. The origin of which is competing interactions due to the intrinsic geometrical configurations like triangular antiferromagnets (geometrical frustration)~\cite{M.Collins1997,R.Moessner2001}. The other interactions such as further-nearest neighbour interaction in $J_1$-$J_2$ systems~\cite{M.Pregelj2015,R.Nath2009}, Dzyaloshinskii-Moriya interaction~\cite{M.Halg2014,W.Jin2017,J.Chung2013}, and single-ion anisotropy~\cite{A.Ramirez1999,S.Isakov2005} can also play a role.

Frustration naturally leads to degenerate ground states that are the source of unusual physical phenomena. For example, a spin ice state emerges in pyrochlore system Dy$_2$Ti$_2$O$_7$~\cite{A.Ramirez1999} due to the degeneracy of ground states with the two-in/two-out spin configuration in a tetrahedral lattice. For Heisenberg triangular antiferromagnets, the ground state has a 120$\,^{\circ}\mathrm{}$ spin structure, which is infinitely degenerate due to the SO(3) symmetry~\cite{H.Kawamura1985}.
As demonstrated many times in the field~\cite{A.Ramirez1999,Y.Shirata2012,J.Gardner2001,K.Ueda2016,E.Ma2015}, finding a new frustrated system is an essential step that leads to discoveries of novel hitherto unreported phenomena.

Rare-earth compounds can be a fertile ground in search for a new frustrated system. The 4f elements provide a strong spin-orbit coupling (SOC), which induces large single-ion anisotropy. Its energy scale can be comparable to other interactions such as the dipole interaction and the exchange interaction energy~\cite{A.Ramirez1999,S.Isakov2005}, leading to competition. The competing interactions with a large SOC that depends on rare-earth elements is the source of various correlated phenomena: spin ice~\cite{A.Ramirez1999}, spin liquid~\cite{J.Gardner2001}, metal-insulator transitions~\cite{K.Ueda2016}, and Weyl semimetals~\cite{E.Ma2015}.

Moreover, the large SOC in rare-earth systems induces $\mathcal{J}$ multiplets as ground states. With a particular crystal electric field (CEF), the states can be further split to make effective $S$=1/2~\cite{A.Ramirez1999,Y.Jana2002,S.Curnoe2013} with strong quantum fluctuations. Phenomena like quantum spin ice~\cite{K.Kimura2013,L.Pan2016} and quantum spin liquid~\cite{J.Gardner2001,H.Takatsu2011,Y.Tokiwa2014} have been recently reported arising from such strong quantum fluctuations in frustrated systems.

The other important factor is the dimensionality of the system. Low-dimensional structures such as one-dimensional (1D) chains and two-dimensional (2D) triangular lattices, are natural candidates for new interesting ground states. This is because reduced dimensionality prevents the magnetic structure from having a simple collinear configuration and leads to more complicated ground states~\cite{N.Mermin1966,P.Anderson1973}. Equally interesting are the prediction of some exotic phases in three-dimensional (3D) systems~\cite{M.Hermele2004,A.Banerjee2008,O.Sikora2009} and the discovery of 3D frustrated materials in pyrochlore and hyperkagome lattices~\cite{Y.Okamoto2007,M.Lawler2008,P.Khuntia2016}.

A honeycomb lattice is a bipartite lattice and thus, it is not, a priori, frustrated with the nearest neighbor interaction alone. However, it can become frustrated with the addition of further nearest neighbor interactions~\cite{R.Bishop2015,J.Rehn2016}. Only few inorganic honeycomb systems with large inter-layer interaction have been so far reported. Bi$_3$Mn$_4$O$_{12}$(NO$_3$) with an antiferromagnetic Heisenberg honeycomb bilayer lattice shows frustration due to the competition between the nearest and next-nearest neighbors~\cite{O.Smirnova2009,H.Kandpal2011}. On the other hand, in Sr$Ln_2$O$_4$ ($Ln$=Gd, Dy, Ho, Er, Tm and Yb), the zigzag ladders of $Ln^{3+}$ running along the $c$ axis are linked to each other and form a honeycomb lattice on the $ab$ plane~\cite{H.Karunadasa2005}. Frustration here leads to an exotic magnetic structure, where both long- and short-range magnetic order coexist at high magnetic field~\cite{T.Hayes2011,D.Quintero-Castro2012}.

In the previous study of CuNd$_2$Ge$_2$O$_8$~\cite{J.Campa1995}, it was found that the NdO$_8$ polyhedra form triangulated dodecahedra. The deviation of the reciprocal susceptibility from the Curie-Weiss law was explained in terms of crystal-field couplings instead of the interaction between Cu$^{2+}$ and Nd$^{3+}$. The unusual structure based on NdO$_8$ polyhedra as well as the absence of a long-range order despite the large Curie-Weiss temperature implies significant frustration present in this Nd compound. However, a clear understanding of the microscopic mechanism remains poorly understood. In this study, we have characterized the crystal structure precisely, in particular with respect to the magnetic ions and determined the comprehensive physical properties of several other Cu$RE_2$Ge$_2$O$_8$ materials ($RE$=Pr, Nd, Sm, Eu). These results provide clear evidence of substantial frustration present in these new antiferromagnetic compounds.

\section{II. EXPERIMENTAL METHODS}

\begin{figure}[t]
\includegraphics[width=\columnwidth,clip]{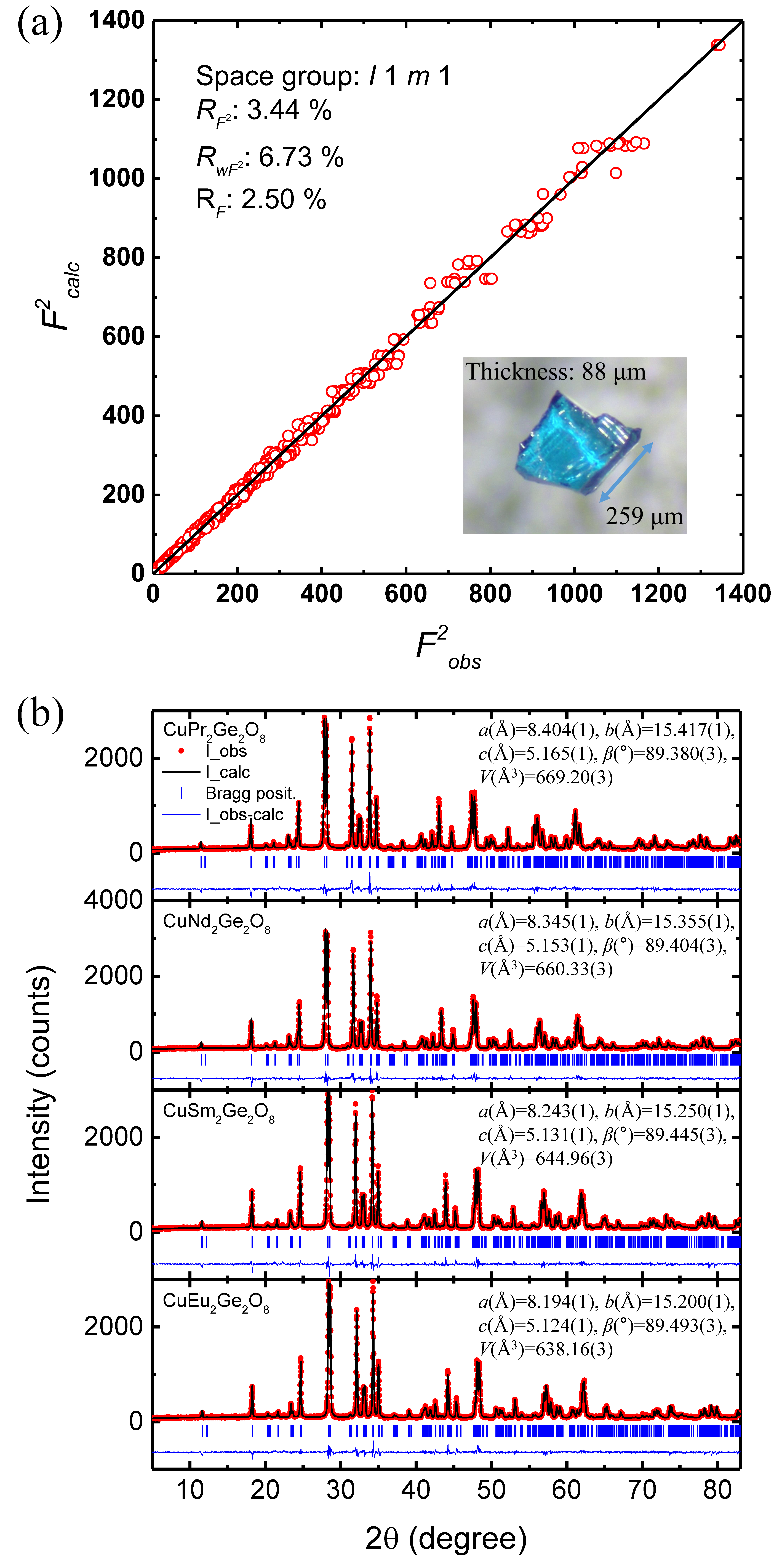}
\caption{ \label{fig1}(Color online) x-ray diffraction data of Cu$RE_2$Ge$_2$O$_8$ ($RE$=Pr, Nd, Sm, Eu) (a) The refinement result of single-crystal diffraction data of CuNd$_2$Ge$_2$O$_8$ and a picture of the single-crystal sample (in the inset). (b) The refinement results of powder diffraction data of Cu$RE_2$Ge$_2$O$_8$. The vertical bars just below the data indicate the position of Bragg peaks. The lines at bottom represent the difference curves.}
\end{figure}

Single crystals of CuNd$_2$Ge$_2$O$_8$ were grown through a self-flux method and several polycrystalline samples of Cu$RE_2$Ge$_2$O$_8$ ($RE$=Pr, Nd, Sm, Eu) were synthesized using a solid state reaction method~\cite{J.Campa1995}. To grow the single crystal of CuNd$_2$Ge$_2$O$_8$, we used CuO and GeO$_2$ powder as self-flux and the ratio of the initial materials was CuO : Nd$_2$O$_3$ : GeO$_2$ = 8 : 1 : 6. The mixture was annealed for 30 minutes at $1260\,^{\circ}\mathrm{C}$ in a platinum crucible and then slowly cooled down to $1000\,^{\circ}\mathrm{C}$ at a rate of $2\,^{\circ}\mathrm{C}$/hour. Using diluted HCl ($\sim$17 \%) solution, we removed residual flux and byproducts such as CuO and CuGeO$_3$ from the grown crystals. The typical volume of CuNd$_2$Ge$_2$O$_8$ single crystals is about 0.001 mm$^3$ as shown in the photo of one of the crystals in the inset of Fig.~\ref{fig1}(a). The color of the crystals is transparent greenish blue and they have an albite shape similar to that reported in Ref.~\cite{J.Campa1995}.

We also prepared several high-purity polycrystalline samples of Cu$RE_2$Ge$_2$O$_8$ by pelletizing and annealing the mixture of CuO, $RE_2$O$_3$, and GeO$_2$ in a stoichiometric ratio, 1 : 1 : 2 for 12 hours at $850\,^{\circ}\mathrm{C}$. The annealed powder of  Cu$RE_2$Ge$_2$O$_8$ was further ground and pelletized before another sintering at elevated temperatures from 950 to $1100\,^{\circ}\mathrm{C}$ at $50\,^{\circ}\mathrm{C}$ steps with a duration of 24 hours at each step. We optimized the final sintering temperature to synthesize the highest-quality samples by carefully monitoring x-ray diffraction (XRD) data. 

\begin{figure*}
\centering
\includegraphics[width=\textwidth,clip]{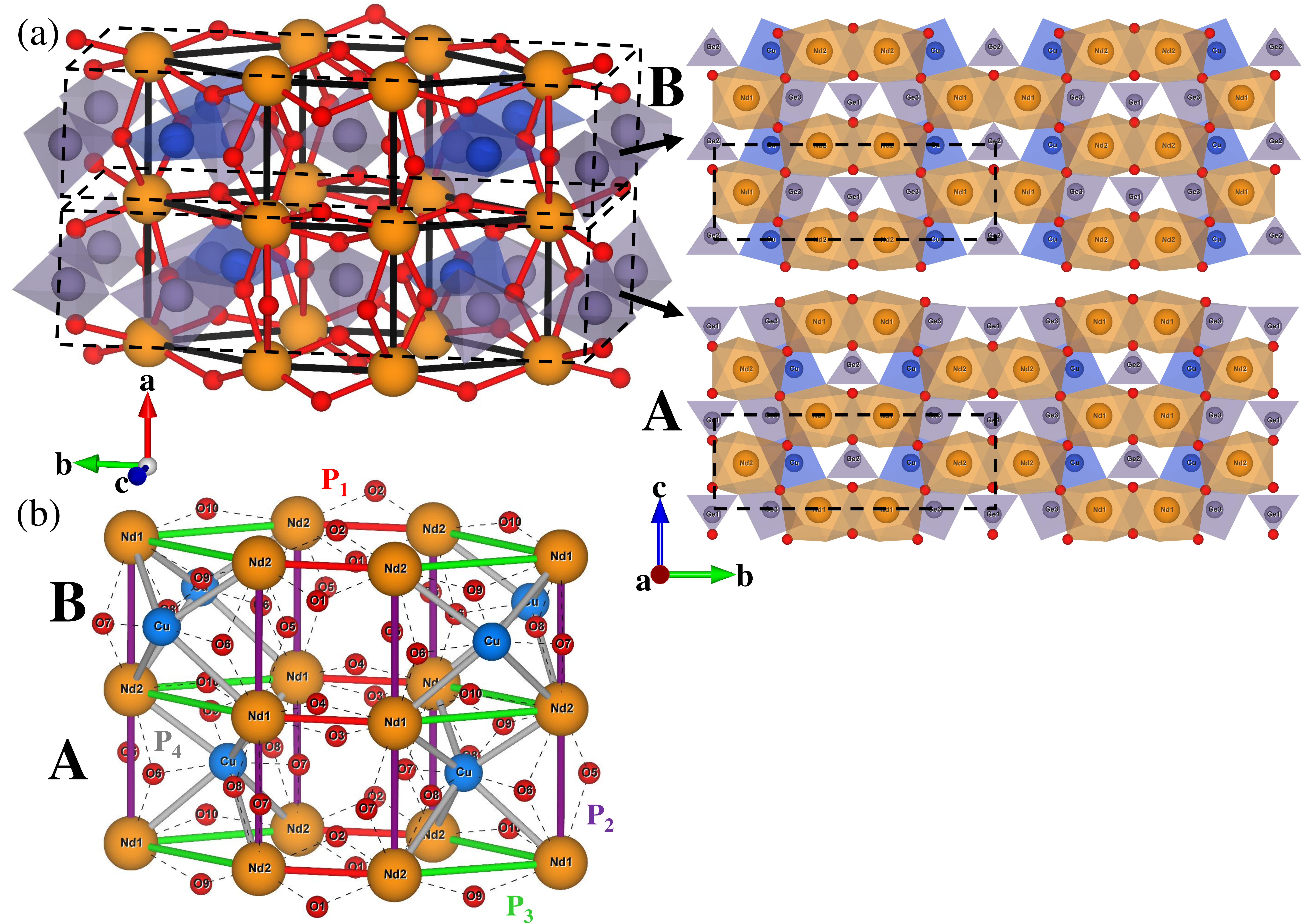}
\caption{\label{fig2}Crystal structure of CuNd$_2$Ge$_2$O$_8$. (a) (Left) The unit cell of CuNd$_2$Ge$_2$O$_8$. The orange, blue, grey, and red balls indicate the neodymium, copper, germanium, and oxygen atoms, respectively. We omit some of the oxygen atoms at the corner of the CuO$_4$ plaquettes, GeO$_4$ tetrahedra and GeO$_5$ bipyramids for simplicity. The red lines display the Nd-O-Nd bond between adjacent Nd atoms, which forms a honeycomb lattice linked along the $a$ axis. (Right) The unit cell can be separated into two sublattices, A and B. The sublattice A contains two copper atoms inside the honeycomb lattice. On the other hand, there are two germanium atoms instead of copper in the sublattice B. (b) The unit cell made of Cu and Nd atoms and the possible exchange paths between the neighboring magnetic atoms: P$_1$, P$_2$, P$_3$, and P$_4$.}
\end{figure*}

To check the quality of the samples, we  performed high-resolution single-crystal and powder x-ray diffraction experiments. For the full structural refinement of the CuNd$_2$Ge$_2$O$_8$ single crystals, we used a XtaLAB P200, Rigaku with a wavelength of 0.710747 \AA~(a Mo target, averaged $K_{\alpha}$). For the structural refinement of polycrystalline Cu$RE_2$Ge$_2$O$_8$ ($RE$=Pr, Nd, Sm, Eu), we used a powder diffractometer (Miniflex II, Rigaku) with wavelengths of 1.540590 and 1.544310 \AA~($K_{\alpha1}$ and $K_{\alpha2}$, respectively). The diffraction data were analyzed using the FULLPROF~\cite{J.Rodriguez1993} software.

Magnetic and thermodynamic properties of the polycrystalline samples were characterized from 0.4 to 350 K using a MPMS-XL5 and a PPMS-9ECII (Quantum Design) equipped with $^3$He/$^4$He options. We carried out magnetization measurements up to 140 kOe at 3 K using a PPMS-14 (Quantum Design) with the vibrating sample magnetometer (VSM) option at the National Center for Inter-University Research Facilities (NCIRF) at Seoul National University.

\section{III. EXPERIMENTAL RESULTS AND ANALYSIS}
\subsection{A. Structural analysis}
We refined the full crystal structure of CuNd$_2$Ge$_2$O$_8$ using a small, high-quality single crystal with the dimensions of 259 $\times$ 260 $\times$ 88 $\mu$m$^3$ after taking into account the x-ray absorption effects. We collected altogether 2889 Bragg peaks (1860 independent Bragg peaks). The crystal structure was refined using the $I~1~m~1$ unit cell setting of the nonmagnetic rare-earth analog~\cite{H.Cho2017}. The unit cell size is $a$(\AA)=8.342(3), $b$(\AA)=15.372(4), $c$(\AA)=5.157(1), $\beta$($\,^{\circ}\mathrm{}$)=89.487(9), and $V$ (\AA$^3$)=661.2(3). The final results shown in Fig.~\ref{fig1}(a) indicate good agreement between the measured data and the refined results. For the structural refinement of other rare-earth cases, we used the data obtained from the powder x-ray diffractometer with the aid of the structural information from the single crystal of CuNd$_2$Ge$_2$O$_8$ [see Fig.~\ref{fig1}(b)]. The difference between the observed and calculated intensity from the refined structure is small as shown at the bottom of each figures. 

Fig.~\ref{fig2} shows the refined crystal structure of the CuNd$_2$Ge$_2$O$_8$ system, which is isostructural with other rare-earth systems including the nonmagnetic rare-earth analog Cu(Y/La)$_2$Ge$_2$O$_8$~\cite{H.Cho2017}. The unit cell size and monoclinic $\beta$ angle vary depending on the ionic size of $RE^{3+}$~\cite{U.Lambert1986}. In the previous study of nonmagnetic rare-earth cases~\cite{H.Cho2017}, the magnetism of the system was characterized by the Cu-ion coordination since Cu is the only magnetic element present in the compound. For magnetic rare-earth cases, however, the magnetism is instead expected to be affected significantly by the large magnetic moments of rare-earth ions.

\begin{table}[h]
\caption{\label{table1}Table of bond lengths and angles along each of exchange paths P$_i$.}
\centering
\renewcommand{\arraystretch}{2}
\begin{tabular}{c c c c c}
\hline
\hline
 \multicolumn{3}{c}{Path} & Bond length ({\AA}) & Bond angle ($\,^{\circ}\mathrm{}$)\\
 \hline
 \multirow{4}{*}{P$_1$} & & Nd1-O3-Nd1 & 4.747(1) & 102.6(2) \\ 
& & Nd1-O4-Nd1 & 4.685(1) & 104.5(3) \\ 
& & Nd2-O1-Nd2 & 4.671(1) & 104.1(3) \\
& & Nd2-O2-Nd2 & 4.729(1) & 102.3(3) \\
 \hline
 \multirow{4}{*}{P$_2$} & & Nd1-O5-Nd2 & 5.495(1) & 99.3(2) \\ 
& & Nd1-O6-Nd2 & 4.843(1) & 119.7(2) \\ 
& & Nd2-O7-Nd1 & 5.514(1) & 97.7(2) \\
& & Nd2-O8-Nd1 & 4.735(1) & 122.6(2) \\
 \hline
 \multirow{2}{*}{P$_3$} & & Nd2-O9-Nd1 & 5.025(1) & 141.2(2) \\ 
& & Nd1-O10-Nd2 & 4.943(1) & 149.2(2) \\ 
 \hline
 \multirow{8}{*}{P$_4$} & & Nd1-O6-Cu & 4.498(1) & 102.6(2) \\ 
& & Nd2-O6-Cu & 4.464(1) & 92.4(2) \\ 
& & Nd1-O7-Cu & 4.635(1) & 94.2(2) \\
& & Nd2-O7-Cu & 4.785(1) & 101.8(2) \\
& & Nd1-O8-Cu & 4.288(1) & 106.6(2) \\ 
& & Nd2-O8-Cu & 4.207(1) & 122.2(3) \\
& & Nd1-O9-Cu & 4.489(1) & 99.0(2) \\
& & Nd2-O9-Cu & 4.482(1) & 91.4(2) \\
\hline
\hline
 \end{tabular}
\end{table}

$RE$O$_8$ and CuO$_4$ form distorted triangular dodecahedra and plaquettes, respectively, while  germanium oxides have GeO$_4$ tetrahedra and GeO$_5$ bipyramids. In the unit cell of CuNd$_2$Ge$_2$O$_8$~[see Fig.~\ref{fig2}(a) (left)], linked by O atoms, Nd atoms form an elongated honeycomb layer on the $bc$ plane and the layers are linked along the $a$ axis. The unit cell can be separated into two sublattices: A and B. The sublattice A contains two Cu atoms inside the honeycomb. On the other hand, sublattice B contains two Ge atoms inside the honeycomb instead of Cu and the Cu atoms are positioned outside the honeycomb unit [see Fig.~\ref{fig2}(a) (right)].  Fig.~\ref{fig2}(b) shows several exchange paths between adjacent magnetic atoms in the sublattices A and B. The paths P$_1$ and P$_2$ are formed by two edge-sharing NdO$_8$ along the $b$ and $a$ axes, respectively. P$_3$ follows paths made by two corner-sharing NdO$_8$ and P$_4$ indicates the bonding of Nd-O-Cu. The bond lengths and angles of Nd-O-Nd and Nd-O-Cu along each path P$_i$ vary in certain ranges as summarized in TABLE~\ref{table1}.

From the coordination of the $RE$ atoms one can immediately see that the magnetic lattice has a honeycomb tunnel-like structure running along the $a$ axis. The interlayer bond length P$_2$ is comparable to the bond lengths in the honeycomb lattice P$_1$ and P$_3$. It suggests that the inter- and intralayer interactions have similar energy scales, which is different from a common 2D honeycomb lattice~\cite{T.Sato2003}. Although the bond lengths of P$_1$, P$_2$, and P$_3$ are roughly similar to one another, the bond angles between the magnetic atoms differ significantly from each other. It implies that the sign of exchange interaction along each path is expected to be different as well~\cite{J.Goodenough1963}. Such a complicated mixture of competing interactions is most likely to be a leading factor to make this system frustrated~\cite{C.Balz2016,S.Derakhshan2008,D.Peets2017,R.Nirmala2017}.

We further note that the structure of Cu$RE_2$Ge$_2$O$_8$ is unique as compared with the previously reported honeycomb systems having an inter-layer interaction. For instance, Bi$_3$Mn$_4$O$_{12}$(NO$_3$) has a honeycomb bilayer lattice made of MnO$_6$ octahedra with a small anisotropy~\cite{O.Smirnova2009,H.Kandpal2011} while Sr$Ln_2$O$_4$ is governed by dominant interactions on zigzag ladders made of $Ln$$O_6$ octahedra~\cite{H.Karunadasa2005,J.Wen2015}. However, the honeycomb lattice of Cu$RE_2$Ge$_2$O$_8$ is made of $RE$O$_8$ triangular dodecahedra, which induce different ground states and anisotropy due to the distinct CEF.

\subsection{B. Magnetic measurements}

\begin{figure}[t]
\includegraphics[width=\columnwidth,clip]{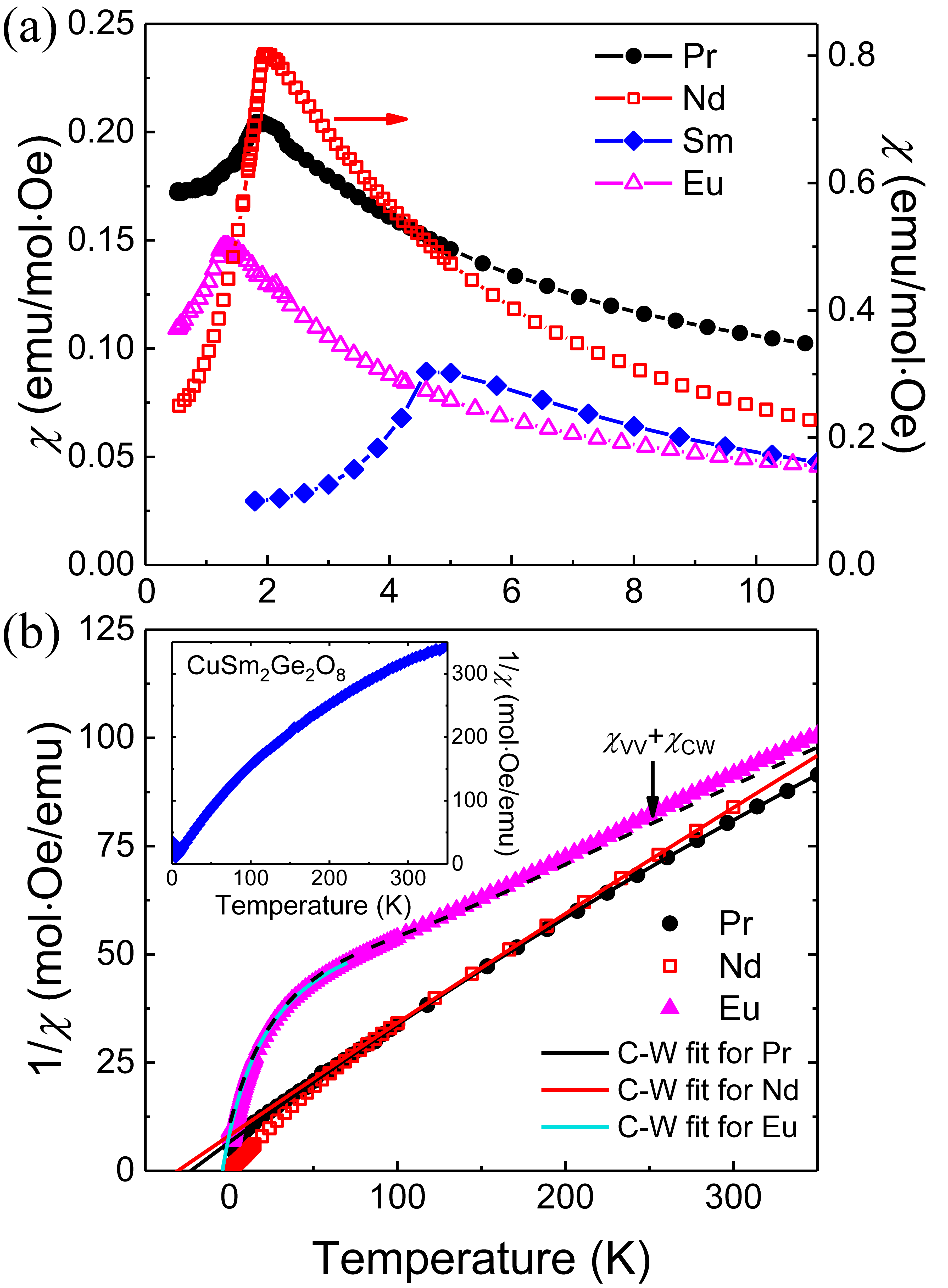}
\caption{\label{fig3}(Color online) (a) Temperature-dependent dc magnetic susceptibility $\chi$($T$) of Cu$RE_2$Ge$_2$O$_8$ ($RE$=Pr, Nd, Sm, Eu). (b) The inverse susceptibility curves and the Curie-Weiss (C-W) fitting to the respective data are shown as solid lines. The inset shows the inverse susceptibility of CuSm$_2$Ge$_2$O$_8$.}
\end{figure}

Fig.~\ref{fig3}(a) shows the temperature-dependent magnetic susceptibility, $\chi(T)$, of Cu$RE_2$Ge$_2$O$_8$. All the systems show a peak below 5 K, which implies an onset of long-range order. We estimate the ordering temperature from the maximum in the d$\chi$/d$T$ curve. Fig.~\ref{fig3}(b) shows the inverse magnetic susceptibility, 1/$\chi$. By applying the Curie-Weiss law,
\begin{equation}
\chi(T) = \chi_0 + C/(T-\theta_{CW})
\label{CWfit}
\end{equation}
we estimated the temperature-independent term $\chi_0$, effective moments of $RE^{3+}$ $\mu_{eff,exp}$, and the Cure-Weiss temperature $\theta_{CW}$ except for the Sm case. The parameters $\mu_{eff,exp}$ were calculated by the following relation:
\begin{equation}
\mu_{eff,exp} = \sqrt{[3(\mu_{eff,total})^2 - (\mu_{eff,Cu})^2]/2}
\label{mueff}
\end{equation}
(here, $\mu_{eff,total}$ is the total effective moment from the Curie constant $C$ and $\mu_{eff,Cu}$ is the effective moment of Cu~\cite{H.Cho2017}). For our fitting of the Pr and Nd compounds, we used the data taken above 100 K. The respective ordering temperatures, the experimental and reference effective moments of $RE^{3+}$~\cite{S.Blundell2001}, and the Curie-Weiss temperatures are summarized in TABLE~\ref{table2}.

\begin{table}[h]
\caption{\label{table2}The summary of the magnetic susceptibility data.\\(*Effective moment of Cu~\cite{H.Cho2017}.)}
\centering
\renewcommand{\arraystretch}{2}
\begin{tabular}{c c c c c c c}
\hline
\hline
  & $\mathcal{J}$ & $T_N$(K) & $\frac{\mu_{eff,exp}}{\mu_{B}}$ & $\frac{\mu_{eff,theo}}{\mu_{B}}$ & $\theta_{CW}$(K) &$f=\frac{|\theta_{CW}|}{T_N}$\\
 \hline
 Pr & 4 & 1.80(1) & 3.44(2) & 3.58 & -23.0(7) & 12.8(4) \\
 Nd & $\frac{9}{2}$ & 1.90(1) & 3.62(3) & 3.62 & -31(1) & 16.3(5) \\
 Sm & $\frac{5}{2}$ & 4.3(1) & Undefined & 0.85 & Undefined & Undefined\\
 Eu & 0 & 1.21(2) & 1.85(2) & 1.94(1)* & -4.1(3) & 3.4(3) \\
  \hline
  \hline
 \end{tabular}
\end{table}

For the Pr case, $\chi_0$ is estimated to be $\sim$0.8$\times$10$^{-3}$ emu/Pr$\cdot$Oe, which is close to Van Vleck paramagnetism due to the crystal electric field effect in Pr$_2$Ir$_2$O$_7$~\cite{S.Nakatsuji2006}. For the Nd case, we found $\chi_0$$\sim$3$\times$10$^{-4}$ emu/Nd$\cdot$Oe. The measured effective moments of the Pr and Nd cases ($\mu_{eff,exp,Pr}$=3.44(2)$\mu_{B}$/Pr and $\mu_{eff,exp,Nd}$=3.62(3)$\mu_{B}$/Nd)  are comparable to the theoretical values of free ion moments, $\mu_{eff,theo}$. The negative sign of $\theta_{CW}$ indicates that the dominant exchange interaction is antiferromagnetic. The ratio of $\theta_{CW}$ to the N\'eel ordering temperature $T_N$, which is defined as a `frustration factor' $f$, is an experimental measure of the strength of frustration~\cite{A.Ramirez1994}. The large value of $f_{Pr}$=12.8 and $f_{Nd}$=16.2 indicates that these rare-earth systems are highly frustrated.  

The samarium compound also shows a cusp in $\chi$($T$) at $T_N$=4.34 K. However, the inverse susceptibility does not follow the Curie-Weiss law due to the intermultiplet splittings with different $\mathcal{J}$~\cite{S.Singh2008,J.pospisil2010} and the experimental effective moments around room temperature do not coincide with the theoretical value from Hund's rules~\cite{S.Blundell2001}. Therefore, we could not estimate the exchange interaction of the ground states from $\theta_{CW}$ obtained by fitting the high temperature region. Moreover, the relatively large $T_N$ practically excludes data fitting in the low-temperature region.

For the Eu compound, the fitting was applied on the data taken in the low-temperature region (21$\sim$70 K) due to the low-lying excited states. $\chi_0$ is found to be $\sim$10$^{-2}$ emu/mol$\cdot$Oe, which is close to Van Vleck paramagnetism of the nonmagnetic ground state $^7$F$_0$ of Eu$_2$O$_3$~\cite{A.S.Borovik1956}. Also, the effective moment $\mu_{eff,exp,Eu}$=1.85(2)$\mu_{B}$/mol is close to the effective moment of Cu$^{2+}$ which suggests that the ground state of the Eu ion in CuEu$_2$Ge$_2$O$_8$ is nonmagnetic. To estimate the overall temperature-dependent Van Vleck contribution (the dashed line in Fig.~\ref{fig3}(b)), the Cu$^{2+}$ contribution was removed by subtracting the Curie-Weiss term of the La compound~\cite{H.Cho2017} from the measured data. The final results (57$\sim$350 K) were used for fitting to the Van Vleck paramagnetic susceptibility~\cite{Y.Takikawa2010}. The fit provides a spin-orbit couping constant of $\lambda$$\sim$423 K, which is similar to that of the Eu$_2$Ir$_2$O$_7$ compound~\cite{J.Ishikawa2012}.  
When we consider just the magnetic moment of Cu$^{2+}$, the system can be characterized as a two-dimensional triangular antiferromagnet on the $ac$ plane~\cite{H.Cho2017}. We note that values of $T_N$ and $\theta_{CW}$ are comparable to those of the Y and La cases where $f_{Y}$=2.2$<$$f_{Eu}$=3.4$<$$f_{La}$=6.1.

\begin{figure}[t]
\includegraphics[width=\columnwidth,clip]{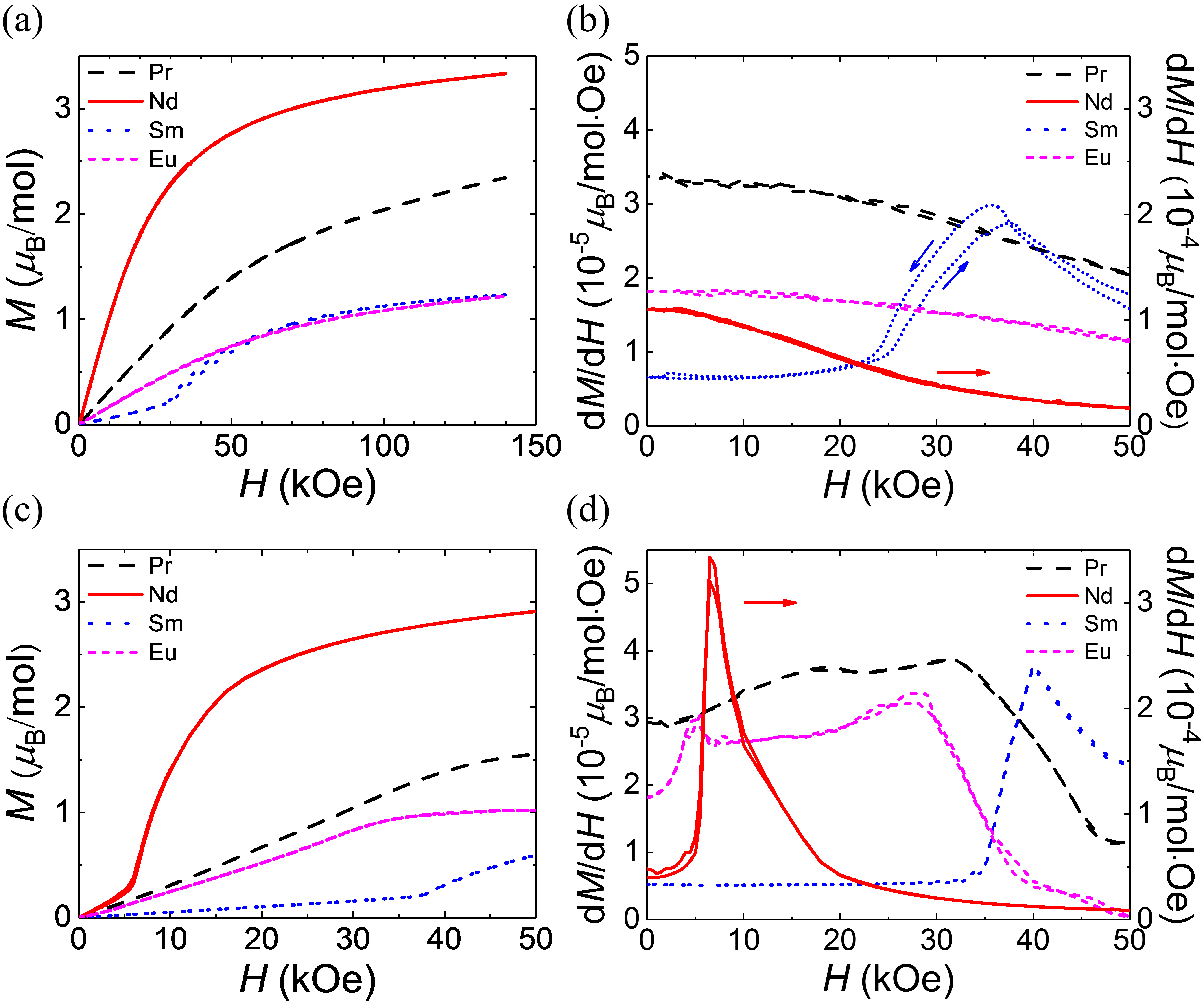}
\caption{\label{fig4}(Color online) (a) Field-dependent magnetization $M$($H$) at 3 K and (b) the first derivative d$M$($H$)/d$H$ at 3 K. (c) Field-dependent magnetization $M$($H$) measured at 0.6 K below $T_N$ and (d) the first derivative d$M$($H$)/d$H$ at 0.6 K.}
\end{figure}

To estimate the ground state and anisotropy of all the samples, we measured the field-dependent magnetization $M$($H$). Fig.~\ref{fig4}(a) and (b) present $M$($H$) taken at 3 K up to 140 kOe and its first derivative, respectively. None of the systems show hysteresis around zero field while ramping up and down external magnetic field. Using our high-field magnetization data, we can infer that the moments of the Nd, Sm, and Eu cases are approaching saturation at 140 kOe but the magnetization of the Pr compound increases even at such high magnetic fields. The need for larger external field to saturate Pr moments implies that the single-ion anisotropy and/or exchange interactions between the moments are relatively stronger for the Pr compound than in the other systems.

The saturated magnetization $M_s$ of the Nd case is $\sim$3.3$\mu_{B}$/mol. Since there are two Nd$^{3+}$ ions per formula unit, subtracting the Cu contribution 1$\mu_{B}$/mol from $M_s$ gives us the Nd contribution equal to 1.15$\mu_{B}$/Nd. This value is similar to $M_s$ of Nd$_2$Ir$_2$O$_7$ with Ising magnetic anisotropy~\cite{Z.Tian2016}. The estimated saturation field $H_{sat}$$\cong$$k_B\theta_{CW}/\mu_{B}g_{\mathcal{J}}\langle\mathcal{J}_z\rangle$=190 kOe with $\langle\mathcal{J}_z\rangle$ calculated using the ground state of Nd$_2$Ir$_2$O$_7$~\cite{M.Watahiki2011} is also comparable to our experimental results.

For the Sm case, we found $M_s$$\sim$1.2$\mu_{B}$/mol. By excluding the Cu$^{2+}$ moments, magnetization coming from Van Vleck paramagnetism and Sm$^{3+}$ ions is found to have only 0.2$\mu_{B}$/mol. Such a small moment of Sm$^{3+}$ is similar to that of frustrated Sm$_2$Ti$_2$O$_7$ system with $\mu_{eff}=0.15\mu_{B}$/Sm~\cite{S.Singh2008}. On the other hand, $M_s$ of the Eu case is $\sim$1.22$\mu_{B}$/mol, which corresponds to the sum of the saturated moment of Cu$^{2+}$ 1$\mu_{B}$/mol and the Van Vleck paramagnetism 0.25$\mu_{B}$/mol. 

Fig.~\ref{fig4}(c) shows $M$($H$) at 0.6 K for all the systems up to 50 kOe and Fig.~\ref{fig4}(d) shows their first derivative d$M$/d$H$ below $T_N$. All the measured $RE$ compounds show a metamagnetic transition below $T_N$ as shown by the d$M$/d$H$ curve. The Pr system reveals two broad humps with a critical field of $H_{c1}$=18 and $H_{c2}$=32 kOe in d$M$/d$H$. We note that the critical fields are different from the saturation field. This indicates field-induced phase transitions, which can be related to the phase transition in the spin configuration of Cu$^{2+}$ and Pr$^{3+}$ moments, respectively. The Nd compound shows a sharp peak at $H_c$=6.5 kOe in d$M$/d$H$. We think this peak is related to an abrupt change in $M$($H$), implying the first-order nature of the transition such as a spin-flop transition. Like the Nd compound, the Sm compound also shows a sharp peak at $H_c$=40 kOe in d$M$/d$H$.

On the other hand, the Eu compound exhibits a d$M$/d$H$ curve (see Fig.~\ref{fig4}(d)) similar to the nonmagnetic rare-earth systems~\cite{H.Cho2017}, with a peak at $H_c$=5.0 kOe and the saturation taking place at $H_{sat}$=29 kOe. The different behaviour of the field-induced transition depending on the rare-earth element indicates that the exchange interactions in the magnetic lattice are significantly affected by the presence of $RE^{3+}$.

\subsection{C. Heat capacity measurements}

\begin{figure}[t]
\includegraphics[width=\columnwidth,clip]{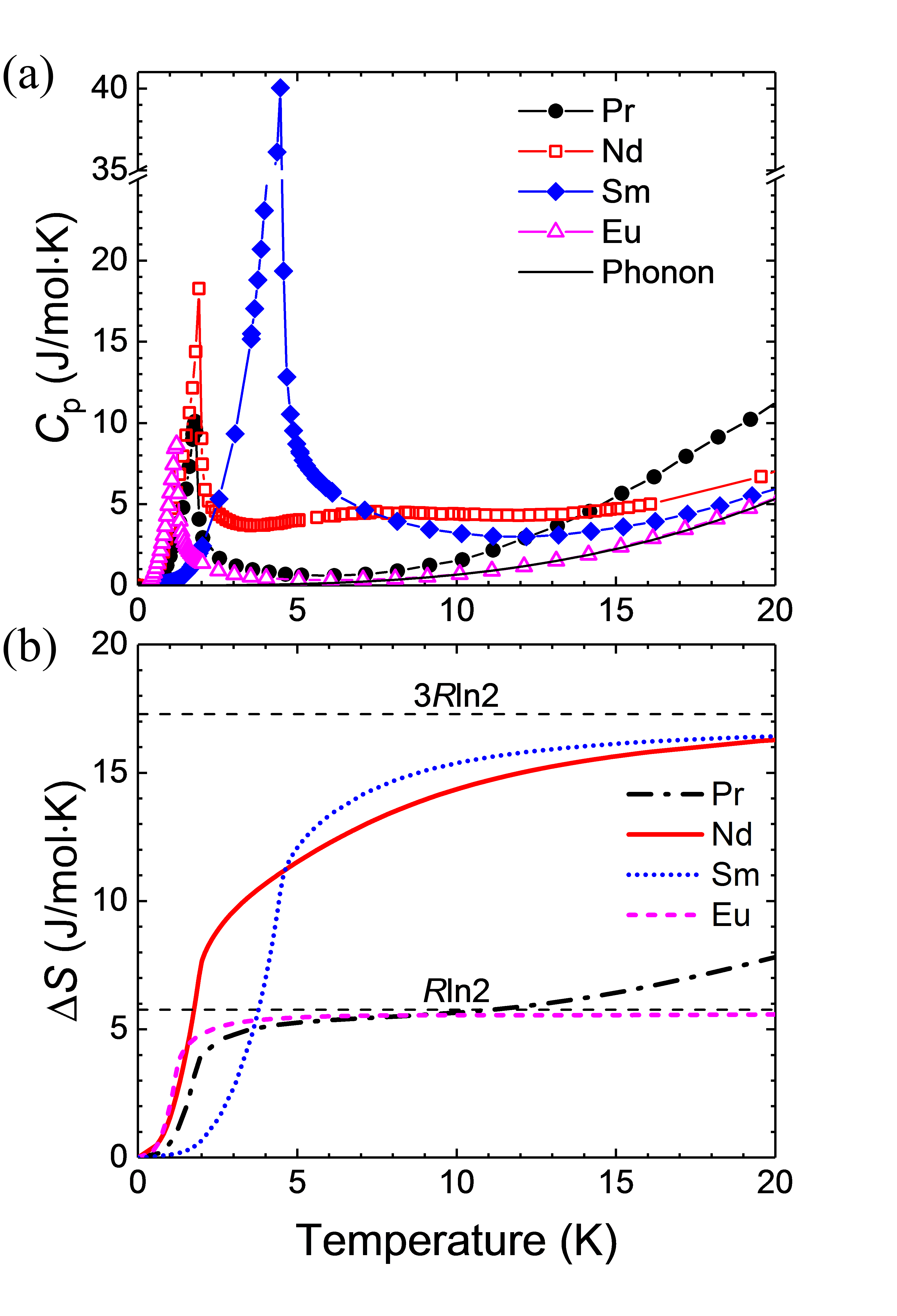}
\caption{\label{fig5}(Color online) (a) Heat capacity $C_p$($T$) of Cu$RE_2$Ge$_2$O$_8$ under zero field. The solid line is the phonon contribution of the Eu case estimated by fitting the $C_p$ to the Debye $T^3$ law. (b) Entropy change $\Delta$$S$ for each $RE$ system. The dashed lines stand for the values of $R$ln2 and 3$R$ln2, respectively.}     
\end{figure}

Fig.~\ref{fig5}(a) shows the temperature dependence of heat capacity $C_p$ of Cu$RE_2$Ge$_2$O$_8$ at zero field. A lambda-like sharp peak observed in all four systems below 5 K confirms an onset of the long-range order as observed in $\chi$($T$). The peak position for every measured $C_p$ curve coincides well with $T_N$ extracted from $\chi$ within the error bar as shown in TABLE~\ref{table2}. Fig.~\ref{fig5}(b) plots the entropy change $\Delta$$S$ of each case. The phonon contribution $C_{phonon}$ of the Eu compound was evaluated using the Debye $T^3$ law in the low-temperature region ($T<20$ K). The $C_{phonon}$ contribution of the other magnetic $RE$ compounds was estimated by scaling the $C_{phonon}$ of Eu~\cite{V.Goruganti2008}. 

For the Eu case, since the spin degree of freedom comes mainly from Cu$^{2+}$ $S$=1/2 (as shown by the magnetization measurements), the entropy change saturates at $R$ln2=5.76 J/mol$\cdot$K. At $T_N$=1.21 K, however, the entropy change $\Delta$$S_{Eu}$ is $\sim$3.39 J/mol$\cdot$K, which corresponds to 59 \% of $R$ln2. The residual entropy released above $T_N$ originates from the correlation that remains short-ranged due to the geometrical frustration of the 2D triangular lattice~\cite{K.Yokota2014,J.Park2010}. 

In CuPr$_2$Ge$_2$O$_8$, the Pr$^{3+}$ ground state can be either a non-Kramers doublet or a nonmagnetic singlet~\cite{S.Nakatsuji2006,A.Princep2013}. Given the value of $\mu_{eff}$ and the significant difference in $M$($H$) between Pr and nonmagnetic rare-earth Eu cases, the ground state of Pr ions in CuPr$_2$Ge$_2$O$_8$ is likely to be the magnetic non-Kramers doublet. In such a case, the total entropy is expected to reach 3$R$ln2 at high temperature including the Cu contribution of $R$ln2. However, the actual entropy obtained at $T_N$ is $\Delta$$S_{Pr}$ is $\sim$3.35 J/mol$\cdot$K at $T_N$=1.8 K, which is only 19 \% of 3$R$ln2 as shown in Fig.~\ref{fig5}(b). Such a reduction of entropy indicates that the spin moments of Pr are highly frustrated similar to the frustrated pyrochlore system $Ln_2$Ir$_2$O$_7$~\cite{K.Matsuhira2011}. The residual entropy is released well above $T>\lvert\theta_{CW}\lvert$ eventually reaching 3$R$ln2 at high temperature. The difference between $C_p$ and $C_{phonon}$ above $T_N$ shows the short-range ordering of the Pr magnetic moments.

In the case of the CuNd$_2$Ge$_2$O$_8$ compound, magnetism at $T_N$ mainly comes from the ground state of the Kramers doublet of Nd; the excited states do not contribute due to the large CEF splittings between the ground and excited states~\cite{M.Watahiki2011}. Therefore, the total entropy change is expected to approach the value of 3$R$ln2 including Cu contribution, consistent with our data in Fig.~\ref{fig5}(b). At $T_N$=1.90 K, $\Delta$$S_{Nd}$$\sim$7.04 J/mol$\cdot$K which corresponds to 0.41$\times$3$R$ln2. Such a small entropy fraction below $T_N$ and large residual entropy above $T_N$ with a broad hump in $C_p$ reflects short-range order in this frustrated system. 

For the Sm compound, we observe a tail above $T_N$ which is a typical sign of short-range order of frustrated systems in the specific heat. Similar to the Nd compound, the ground state of Sm$^{3+}$ is a Kramers doublet. Again, the respective value of $\Delta$$S_{Sm}$$\sim$10.37 J/mol$\cdot$K is observed as shown in Fig.~\ref{fig5}(b) at $T_N$=4.3 K. The entropy difference reaches 0.60$\times$3$R$ln2. This small entropy fraction and the large residual entropy above $T_N$ is also linked to short-range interactions in this frustrated system. 

\section{IV. DISCUSSION}
Cu$RE_2$Ge$_2$O$_8$ ($RE$=Pr, Nd, Sm, Eu) compounds are candidates for a new frustrated system of rare-earth magnetism. For the Eu case, physical quantities such as $\mu_{eff}$, $M_s$, and $\Delta$$S$ show that the ground state of Eu$^{3+}$ is nonmagnetic. Therefore CuEu$_2$Ge$_2$O$_8$ is a 2D triangular antiferromagnet with Cu$^{2+}$ $S$=1/2 quantum spins lying on the $ac$ plane. The triangular plane corrugates in a zigzag shape along the $a$ axis~\cite{H.Cho2017}. The N\'eel temperature of the Eu compound is comparable to $T_{N,Y}$=0.51 and $T_{N,La}$=1.09 K.

In the case of the Pr and Nd compounds, there is an anomaly in $\chi$($T$) around $T_N$. Above $T>T_N$, it follows paramagnetic behavior with a relatively small temperature-independent Van Vleck paramagnetic term related to large crystal-field splittings from the ground state to higher excited states. The small Van Vleck term suggests that $\theta_{CW}$ is mainly contributed by the exchange interactions and not by CEF~\cite{S.Nakatsuji2006}.  
For a low-dimensional magnetic system, one would expect a broad hump~\cite{J.Bonner1964,M.Hase1993,B.Koteswararao2007,N.kini2006,N.Elstner1993} at $\sim$$\lvert\theta_{CW,Pr}\rvert$=23 K and $\lvert\theta_{CW,Nd}\rvert$=31 K. Absence of this feature supports the 3D antiferromagnetic scenario in these systems rather than 1D or 2D. 

The large frustration factors $f_{Pr}$=12.8 and $f_{Nd}$=16.3 obtained from our magnetic susceptibility data point to the substantial frustration in these two systems. The results of the thermodynamic measurements support the significant frustration scenario as well. Similar to other frustrated systems~\cite{T.Radu2005,B.Canals2016}, the entropy change does not reach 3$R$ln2 corresponding to the spin degree of freedom of the ground state below $T_N$. The broad hump in the $C_p$ curve signals that the residual entropy is released above $T_N$. This is quite different from an unfrustrated system, which releases the majority of its entropy below the ordering temperature~\cite{V.Goruganti2008}. The hump in $C_p$ for a rare-earth system might also be explained by the Schottky anomaly due to the splittings of the ground state $\mathcal{J}$ manifold~\cite{E.Gopal1966}. In this case, the temperature scale of the anomaly is expected to be comparable to the energy level splittings. We note that the small Van Vleck paramagnetism of the Pr and Nd cases implies that the spitting is in the order of 100 K considering the similar $RE$O$_8$ dodecahedra of $RE_2X_2$O$_7$ ($X$=Ir,Sn)~\cite{S.Nakatsuji2006,H.Zhou2008,M.Watahiki2011}. However, the Nd compound shows the broad hump in $C_p-C_{phonon}$ at temperatures well below 100 K, which indicates that the hump originates from frustration rather than a Schottky anomaly.

Along with other $RE$ compounds, the Sm case is also a candidate material for a 3D frustrated system. Like Pr and Nd cases, the absence of a broad hump in $\chi$($T$) indicates 3D magnetic lattice. Even though we cannot determine the frustration factor, the thermodynamic measurement shows the presence of frustrated moments. For example, the tail in $C_p-C_{phonon}$ above $T_N$ appears due to a short-range correlation even surviving at high temperatures. The large fraction of entropy change released above $T_N$ shows a sign of short-range order.

The magnitude of $T_N$ for the Pr case is around 1 K, which is comparable to $T_N$ of their nonmangetic
rare-earth counterparts (the Eu compound in Fig.~\ref{fig5} and the Y/La compounds in~\cite{H.Cho2017}). Moreover, the entropy contribution from the sharp peak is similar to the entropy contribution of Cu$^{2+}$ $R$ln2 rather than 3$R$ln2. These aspects suggest that only Cu$^{2+}$ moments order while the $RE^{3+}$ moments do not. Nevertheless, this issue cannot be resolved without the help of neutron diffraction which is left for a future study.

\section{V. CONCLUSION}
We have synthesized a single crystal of CuNd$_2$Ge$_2$O$_8$ and several polycrystalline samples of Cu$RE_2$Ge$_2$O$_8$ ($RE$=Pr, Nd, Sm, Eu), analyzed their crystal structure, and characterized their magnetic and thermodynamic properties. While the ground state of the Eu compound is driven solely by the Cu ions, the Pr, Nd, and Sm compounds show a 3D honeycomb-tunnel-like lattice made of $RE^{3+}$. They display a large frustration factor, a small fraction of the entropy released at $T_N$ (19$\sim$60 \% of 3$R$ln2), and short-range order above $T_N$, characteristic behavior of a frustrated antiferromagnet. Frustration in Cu$RE_2$Ge$_2$O$_8$ is induced by the interplay of crystal structure providing a new route to bring in competing interactions and the anisotropy from the large spin-orbit coupling of the rare-earth elements.

\section{ACKNOWLEDGEMENTS}
We thank Santu Baidya, Choong Hyun Kim, and Yukio Noda for discussion. We also acknowledge Kohki Satoh from Quantum Design Japan Inc. and Young-Mi Song from NCIRF for their helps with some of the measurements. The work at the IBS CCES was supported by the Institute for Basic Science in Korea (IBS-R009-G1).


\begin{thebibliography}{99}

\bibitem{M.Collins1997} M. F. Collins and O. A. Petrenko, Can. J. Phys. \textbf{75}, 605 (1997).

\bibitem{R.Moessner2001} R. Moessner, Can. J. Phys. \textbf{79}, 1283 (2001).

\bibitem{M.Pregelj2015} M. Pregelj, A. Zorko, O. Zaharko, H. Nojiri, H. Berger, L. C. Chapon, and D. Ar\v con, Nat. Commun. \textbf{6}, 7255 (2015).

\bibitem{R.Nath2009} R. Nath, Y. Furukawa, F. Borsa, E. E. Kaul, M. Baenitz, C. Geibel, and D. C. Johnston, Phys. Rev. B \textbf{80}, 214430 (2009).

\bibitem{M.Halg2014} M. H\"alg, W. E. A. Lorenz, K. Yu. Povarov, M. M\aa nsson, Y. Skourski, and A. Zheludev, Phys. Rev. B \textbf{90}, 174413 (2014).

\bibitem{W.Jin2017} W. Jin and O. A. Starykh, J. Phys.: Conf. Ser. \textbf{828}, 012019 (2017). 

\bibitem{J.Chung2013} Jae-Ho Chung, Young Sang Song, Sungil Park, Hiroaki Ueda, Yutaka Ueda, Seung-Hun Lee, J. Korean Phys. Soc. \textbf{62}, 12, 1900 (2013). 

\bibitem{A.Ramirez1999} A. P. Ramirez, A. Hayashi, R. J. Cava, R. Siddharthan, and B. S. Shastry, Nature \textbf{399}, 333 (1999).

\bibitem{S.Isakov2005} S. V. Isakov, R. Moessner, and S. L. Sondhi, Phys. Rev. Lett. \textbf{95}, 217201 (2005).

\bibitem{H.Kawamura1985} H. Kawamura and S. Miyashita, J. Phys. Soc. Jpn. \textbf{54}, 4530 (1985).

\bibitem{Y.Shirata2012} Y. Shirata, H. Tanaka, A. Matsuo, and K. Kindo, Phys. Rev. Lett. \textbf{108}, 057205 (2012).

\bibitem{J.Gardner2001} J. S. Gardner, B. D. Gaulin, A. J. Berlinsky, P. Waldron, S. R. Dunsiger, N. P. Raju, and J. E. Greedan, Phys. Rev. B \textbf{64}, 224416 (2001).

\bibitem{K.Ueda2016} K. Ueda, J. Fujioka, and Y. Tokura, Phys. Rev. B \textbf{93}, 245120 (2016).

\bibitem{E.Ma2015} Eric Yue Ma, Yong-Tao Cui, Kentaro Ueda, Shujie Tang, Kai Chen, Nobumichi Tamura, Phillip M. Wu, Jun Fujioka, Yoshinori Tokura, Zhi-Xun Shen, Science \textbf{350}, 6260 538 (2015).

\bibitem{Y.Jana2002} Y. M. Jana, A. Sengupta, D. Ghosh, J. Magn. Magn. Mater. \textbf{248}, 7 (2002).

\bibitem{S.Curnoe2013} S. H. Curnoe, Phys. Rev. B \textbf{88}, 014429 (2013).

\bibitem{L.Pan2016} LiDong Pan,	N. J. Laurita,	Kate A. Ross, Bruce D. Gaulin, and N. P. Armitage, Nat. Phys. \textbf{12}, 361 (2016).

\bibitem{K.Kimura2013} K. Kimura, S. Nakatsuji, J-J. Wen, C. Broholm, M. B. Stone, E. Nishibori, and H. Sawa, Nat. Commun. \textbf{4}, 1934 (2013).

\bibitem{H.Takatsu2011} H. Takatsu, H. Kadowaki, T. J. Sato, J. W. Lynn, Y. Tabata, T. Yamazaki, and K. Matsuhira, J. Phys.: Condens. Matter \textbf{24}, 052201 (2012).

\bibitem{Y.Tokiwa2014} Y. Tokiwa, J. J. Ishikawa, S. Nakatsuji, and P. Gegenwart, Nat. Mater. \textbf{13}, 356 (2014).

\bibitem{N.Mermin1966} N. D. Mermin and H. Wagner, Phys. Rev. Lett. \textbf{17}, 1133 (1966).

\bibitem{P.Anderson1973} P. W. Anderson, Mater. Res. Bull. \textbf{8}, 153 (1973).

\bibitem{M.Hermele2004} Michael Hermele, Matthew P. A. Fisher, and Leon Balents, Phys. Rev. B \textbf{69}, 064404 (2004).

\bibitem{A.Banerjee2008} Argha Banerjee, Sergei V. Isakov, Kedar Damle, and Yong Baek Kim, Phys. Rev. Lett. \textbf{100}, 047208 (2008) 

\bibitem{O.Sikora2009} Olga Sikora, Frank Pollmann, Nic Shannon, Karlo Penc, and Peter Fulde, Phys. Rev. Lett. \textbf{103}, 247001 (2009).

\bibitem{Y.Okamoto2007} Yoshihiko Okamoto, Minoru Nohara, Hiroko Aruga-Katori, and Hidenori Takagi, Phys. Rev. Lett. \textbf{99}, 137207 (2007).

\bibitem{M.Lawler2008} Michael J. Lawler, Hae-Young Kee, Yong Baek Kim, and Ashvin Vishwanath, Phys. Rev. Lett. \textbf{100}, 227201 (2008).

\bibitem{P.Khuntia2016} P. Khuntia, F. Bert, P. Mendels, B. Koteswararao, A. V. Mahajan, M. Baenitz, F. C. Chou, C. Baines, A. Amato, and Y. Furukawa, Phys. Rev. Lett. \textbf{116}, 107203 (2016).

\bibitem{R.Bishop2015} R. F. Bishop, P. H. Y. Li, O. G\"otze, J. Richter, and C. E. Campbell, Phys. Rev. B \textbf{92}, 224434 (2015).

\bibitem{J.Rehn2016} J. Rehn, Arnab Sen, Kedar Damle, and R. Moessner, Phys. Rev. Lett. \textbf{117}, 167201 (2016).

\bibitem{O.Smirnova2009} O. Smirnova, M. Azuma, N. Kumada, Y. Kusano, M. Matsuda, Y. Shimakawa, T. Takei , Y. Yonesaki, and N. Kinomura, J. Am.
Chem. Soc. \textbf{131}, 8313 (2009).

\bibitem{H.Kandpal2011} Hem C. Kandpal and Jeroen van den Brink, Phys. Rev. B \textbf{83}, 140412(R) (2011).

\bibitem{H.Karunadasa2005} H. Karunadasa, Q. Huang, B. G. Ueland, J. W. Lynn, P. Schiffer, K. A. Regan, and R. J. Cava, Phys. Rev. B \textbf{71}, 144414 (2005).

\bibitem{T.Hayes2011} T. J. Hayes, G. Balakrishnan, P. P. Deen, P. Manuel, L. C. Chapon, and O. A. Petrenko, Phys. Rev. B \textbf{84}, 174435 (2011).

\bibitem{D.Quintero-Castro2012} D. L. Quintero-Castro, B. Lake, M. Reehuis, A. Niazi, H. Ryll, A. T. M. N. Islam, T. Fennell, S. A. J. Kimber, B. Klemke, J. Ollivier, V. Garcia Sakai, P. P. Deen, and H. Mutka, Phys. Rev. B \textbf{86}, 064203 (2012).

\bibitem{J.Campa1995} J. Camp\'a, E. Guti\'errez-Puebla, M. Monge, C. Valero, J. Mira, J. Rivas, C. Cascales, and I. Rasines, J. Solid State Chem. \textbf{120}, 254 (1995).

\bibitem{J.Rodriguez1993} J. Rodriguez-Carvajal, Physica B \textbf{192}, 55 (1993).

\bibitem{H.Cho2017} Hwanbeom Cho, Marie Kratochv\'ilov\'a, Hasung Sim, Ki-Young Choi, Choong Hyun Kim, Carley Paulsen, Maxim Avdeev, Darren C. Peets, Younghun Jo, Sanghyun Lee, Yukio Noda, Michael J. Lawler, and Je-Geun Park, Phys. Rev. B \textbf{95}, 144404 (2017).

\bibitem{U.Lambert1986} U. Lambert and W. Eysel, Powder Diffraction \textbf{1}, 3, 256 (1986).

\bibitem{T.Sato2003} T. J. Sato, S. -H. Lee, T. Katsufuji, M. Masaki, S. Park, J. R. D. Copley, and H. Takagi, Phys. Rev. B \textbf{68}, 014432 (2003).

\bibitem{J.Goodenough1963} J. Goodenough, \textit{Magnetism and the Chemical Bond} (Interscience, New York, 1963).

\bibitem{C.Balz2016} Christian Balz, Bella Lake, Johannes Reuther, Hubertus Luetkens, Rico Sch\"{o}nemann, Thomas Herrmannsd\"{o}rfer, Yogesh Singh, A. T. M. Nazmul Islam, Elisa M. Wheeler, Jose A. Rodriguez-Rivera, Tatiana Guidi, Giovanna G. Simeoni, Chris Baines, and Hanjo Ryll, Nat. Phys. \textbf{12}, 942 (2016).

\bibitem{S.Derakhshan2008} Shahab Derakhshan, John E. Greedan, Tetsuhiro Katsumata, and Lachlan M. D. Cranswick, Chem. Mater. \textbf{20}, 5714 (2008).

\bibitem{R.Nirmala2017} R Nirmala, Kwang-Hyun Jang, Hasung Sim, Hwanbeom Cho, Junghwan Lee, Nam-Geun Yang, Seongsu Lee, R M Ibberson, K Kakurai, M Matsudau, S-W Cheong, V V Gapontsev, S V Streltsov, and Je-Geun Park, J. Phys.: Condens. Matter. \textbf{29}, 13 (2017).

\bibitem{D.Peets2017} Darren C. Peets, Hasung Sim, Seongil Choi, Maxim Avdeev, Seongsu Lee, Su Jae Kim, Hoju Kang, Docheon Ahn, and Je-Geun Park, Phys. Rev. B \textbf{95}, 014424 (2017).

\bibitem{J.Wen2015} J.-J. Wen, W. Tian, V. O. Garlea, S. M. Koohpayeh, T. M. McQueen, H.-F. Li, J.-Q. Yan, J. A. Rodriguez-Rivera, D. Vaknin, and C. L. Broholm, Phys. Rev. B \textbf{91}, 054424 (2015).

\bibitem{S.Blundell2001} S. Blundell, \textit{Magnetism in Condensed Matter} (Oxford University Press, New York, 2001).

\bibitem{S.Nakatsuji2006} S. Nakatsuji, Y. Machida, Y. Maeno, T. Tayama, T. Sakakibara, J. van Duijn, L. Balicas, J. N. Millican, R. T. Macaluso, and Julia Y. Chan, Phys. Rev. Lett. \textbf{96}, 087204 (2006).

\bibitem{A.Ramirez1994} A. Ramirez, Annu. Rev. Mater. Sci. \textbf{24}, 453 (1994).

\bibitem{S.Singh2008} Surjeet Singh, Surajit Saha, S. K. Dhar, R. Suryanarayanan, A. K. Sood, and A. Revcolevschi, Phys. Rev. B \textbf{77}, 054408 (2008).

\bibitem{J.pospisil2010} J. Posp\'i\v sil, M. Kratochv\'ilov\'a, J. Prokle\v ska, M. Divi\v s, and V. Sechovsk\'y, Phys. Rev. B \textbf{81}, 024413 (2010).

\bibitem{A.S.Borovik1956} A. S. Borovik-Romanov and N. M. Kreines, SOVIET PHYSICS JETP \textbf{2}, 4 (1956).

\bibitem{Y.Takikawa2010} Y. Takikawa, S. Ebisu, and S. Nagata, J. Phys. Chem. Solids \textbf{71}, 1592 (2010).

\bibitem{J.Ishikawa2012} Jun J. Ishikawa, Eoin C. T. O'Farrell, and Satoru Nakatsuji, Phys. Rev. B \textbf{85}, 245109 (2012).

\bibitem{Z.Tian2016} Zhaoming Tian, Yoshimitsu Kohama, Takahiro Tomita, Hiroaki Ishizuka, Timothy H. Hsieh, Jun J. Ishikawa,	Koichi Kindo, Leon Balents, and Satoru Nakatsuji, Nat. Phys. \textbf{12}, 134 (2016).

\bibitem{M.Watahiki2011} M. Watahiki, K. Tomiyasu, K. Matsuhira, K. Iwasa, M. Yokoyama, S. Takagi, M. Wakeshima, and Y. Hinatsu, J. Phys. Conf. Ser. \textbf{320}, 012080 (2011).

\bibitem{V.Goruganti2008} V. Goruganti, K. D. D. Rathnayaka, Joseph H. Ross, Jr., Y. \"Oner, C. S. Lue, and Y. K. Kuo, J. Appl. Phys. \textbf{103}, 073919 (2008).

\bibitem{K.Yokota2014} Kazuya Yokota, Nobuyuki Kurita, and Hidekazu Tanaka, Phys. Rev. B \textbf{90}, 014403 (2014).

\bibitem{J.Park2010} Junghwan Park, Seongsu Lee, Misun Kang, Kwang-Hyun Jang, Changhee Lee, S. V. Streltsov, V. V. Mazurenko, M. V. Valentyuk, J. E. Medvedeva, T. Kamiyama, and J.-G. Park, Phys. Rev. B \textbf{82}, 054428 (2010).

\bibitem{A.Princep2013} A. J. Princep, D. Prabhakaran, A. T. Boothroyd, and D. T. Adroja, Phys. Rev. B \textbf{88}, 104421 (2013)

\bibitem{K.Matsuhira2011} Kazuyuki Matsuhira, Makoto Wakeshima, Yukio Hinatsu, and Seishi Takagi, J. Phys. Soc. Jpn. \textbf{80}, 094701 (2011).

\bibitem{J.Bonner1964} J. Bonner and M. Fisher, Phys. Rev. \textbf{135}, A640 (1964).

\bibitem{M.Hase1993} Masashi Hase, Ichiro Terasaki, and Kunimitsu Uchinokura, Phys. Rev. Lett. \textbf{70}, 3651 (1993).

\bibitem{B.Koteswararao2007} B. Koteswararao, S. Salunke, A. V. Mahajan, I. Dasgupta, and J. Bobroff, Phys. Rev. B \textbf{76}, 052402 (2007)

\bibitem{N.kini2006} N. Kini, E. Kaul, and C. Geibel, J. Phys.: Condens. Matter \textbf{18},
1303 (2006).

\bibitem{N.Elstner1993} N. Elstner, R. R. P. Singh, and A. P. Young, Phys. Rev. Lett. \textbf{71}, 1629 (1993).

\bibitem{T.Radu2005} T. Radu, H. Wilhelm, V. Yushankhai, D. Kovrizhin, R. Coldea, Z. Tylczynski, T. Lühmann, and F. Steglich, Phys. Rev. Lett. \textbf{95}, 127202 (2005).

\bibitem{B.Canals2016} Benjamin Canals, Ioan-Augustin Chioar, Van-Dai Nguyen, Michel Hehn, Daniel Lacour, Fran\c{c}ois Montaigne, Andrea Locatelli, Tevfik Onur Mente\c{s}, Benito Santos Burgos, and Nicolas Rougemaille, Nat. Commun. \textbf{7}, 11446 (2016).

\bibitem{E.Gopal1966} E. S. R. Gopal, \textit{Specific Heats at Low Temperatures} (Plenum Press, New York, 1966).

\bibitem{H.Zhou2008} H. D. Zhou, C. R. Wiebe, J. A. Janik, L. Balicas, Y. J. Yo, Y. Qiu, J. R. D. Copley, and J. S. Gardner, Phys. Rev. Lett. \textbf{101}, 227204 (2008).


\end{thebibliography}
\end{document}